\documentclass[aps,prl,twocolumn,groupedaddress]{revtex4-2}
\usepackage{latexsym}
\usepackage{amsmath}
\usepackage{amsfonts}
\usepackage{amssymb}
\usepackage{amscd}
\usepackage{bbm}
\usepackage{bbold}
\usepackage{tikz}
\usepackage{tikz-cd}
\usetikzlibrary{shapes}
\usepackage{hyperref}
\usepackage[export]{adjustbox}
\usepackage[caption=false]{subfig}
\usepackage{soul}

\newcommand{\ket}[1]{|#1\rangle}
\newcommand{\bra}[1]{\langle#1|}
\newcommand{\bracket}[2]{\langle#1|#2\rangle}
\newcommand{\vev}[1]{\langle#1\rangle}

\renewcommand{\flat}{\textbf{1}}
\newcommand{\tr}[1]{{\rm Tr} \left[ #1 \right]}

\newcommand{\heading}[1]{\vspace{0.25truecm}\emph{#1.--}}

\begin{document}
	
\title{Emergent information dynamics in many-body interconnected systems}

\author{Wout Merbis}
\email[]{w.merbis@uva.nl}
\affiliation{Dutch Institute for Emergent Phenomena (DIEP), Institute for Theoretical Physics (ITFA), University of Amsterdam, Science Park 904, 1098 XH Amsterdam, The Netherlands}

\author{Manlio de Domenico}
\email[]{manlio.dedomenico@unipd.it}
\affiliation{Department of Physics and Astronomy ``Galileo Galilei", University of Padua, Via F. Marzolo 8, 315126 Padova, Italy}
\affiliation{Istituto Nazionale di Fisica Nucleare, Sez. Padova, Italy}
 
\date{\today}

\begin{abstract}
	The information implicitly represented in the state of physical systems allows one to analyze them with analytical techniques from statistical mechanics and information theory. In the case of complex networks such techniques are inspired by quantum statistical physics and have been used to analyze biophysical systems, from virus-host protein-protein interactions to whole-brain models of humans in health and disease. Here, instead of node-node interactions, we focus on the flow of information between network configurations. Our numerical results unravel fundamental differences between widely used spin models on networks, such as voter and kinetic dynamics, which cannot be found from classical node-based analysis. Our model opens the door to adapting powerful analytical methods from quantum many-body systems to study the interplay between structure and dynamics in interconnected systems.
\end{abstract}

\maketitle

Complex systems are characterized by emergent phenomena, which cannot be reduced to the individual constituents, but are the result of collective behavior due to their interaction among each other. Consequently, intense research activity has been devoted to better understand network structure, dynamics and their interplay \cite{watts1998collective,barabasi1999emergence,albert2002statistical,caldarelli2002scale,reichardt2004detecting,song2005self,serrano2008self,burda2009localization,de2013mathematical,brockmann2013hidden,hens2019spatiotemporal}, as well as critical phenomena emerging from non-trivial connectivity patterns~\cite{dorogovtsev2008critical,buldyrev2010catastrophic,rozenfeld2010small,bradde2010critical,liu2012core}. 
A distinctive feature of biophysical and socio-ecological systems is that their units exchange and integrate information through the underlying connections, while simultaneously processing this information~\cite{crutchfield1995evolution,crutchfield2012between}: overall, the result leads to complex organization and functionality~\cite{bascompte2003nested,guimera2005functional,bullmore2012economy}. A theory of information flow through unit-unit interactions has been introduced~\cite{de2016spectral} and developed~\cite{ghavasieh2020statistical,ghavasieh2022statistical,villegas2022rg,baccini2022weighted} to incorporate information-theoretic measures -- inspired by quantum statistical mechanics -- that have found applications from characterizing virus-host interactions in the human cell~\cite{ghavasieh2021multiscale} to information exchange in whole-brain models of the human brain in healthy and pathological conditions~\cite{nicolini2020scale,benigni2021persistence,Villegas2022laplacian}.

While being able to describe a wide range of dynamical processes on networks that can be approximated by linearized dynamics, such as diffusive processes and synchronization close to a metastable state~\cite{de2017diffusion}, this framework is constrained to systems where the information field on each node is characterized by a scalar quantity and where its flow is constrained to node-node interactions. In many empirical settings of physical interest, the nodes could carry multiple types of information, while node-based dynamics might be governed by non-linear differential equations. A relevant example where these conditions are met is in reaction-diffusion systems~\cite{hethcote2000mathematics,colizza2007reaction,belik2011natural,PastorSatorras2015,merbis2022logistic}, where nodes are characterized by several different states (compartments) and dynamical rules depending on those states are typically described by a set of coupled non-linear differential equations~\cite{pastor2001epidemic,may2001infection,van2008virus,van2011n}.

In this Letter, we propose a statistical field theory of information dynamics between microstates, i.e. in the space of network configurations, and show that this choice overcomes the limitations of a node-based approach. In fact, in our framework the nodes can have vector-valued information fields and their dynamics reflects information flow between different network states. We show that the classical node-based dynamics can be recovered by suitable projections on the node subspace. Remarkably, our framework is capable of numerically tracking correlations (and higher moments) of node-based observables exactly, a feature that has no counterpart in the state of the art: we show how this is relevant by analyzing two dynamical systems -- the voter model and the (symmetric) simple exclusion process (SEP)~\cite{holley1975ergodic,liggett1985interacting,liggett1997stochastic} -- whose node-based dynamics is governed by the same propagator -- i.e., the graph Laplacian -- but the dynamics on the space of network configurations is wildly different. 
In a joint paper \cite{predraft}, we study the application to network epidemiology, as an example of a stochastic process with a dynamical phase transition.

\heading{Model} 
\begin{figure*}[!ht]
	\includegraphics[width=0.8\textwidth]{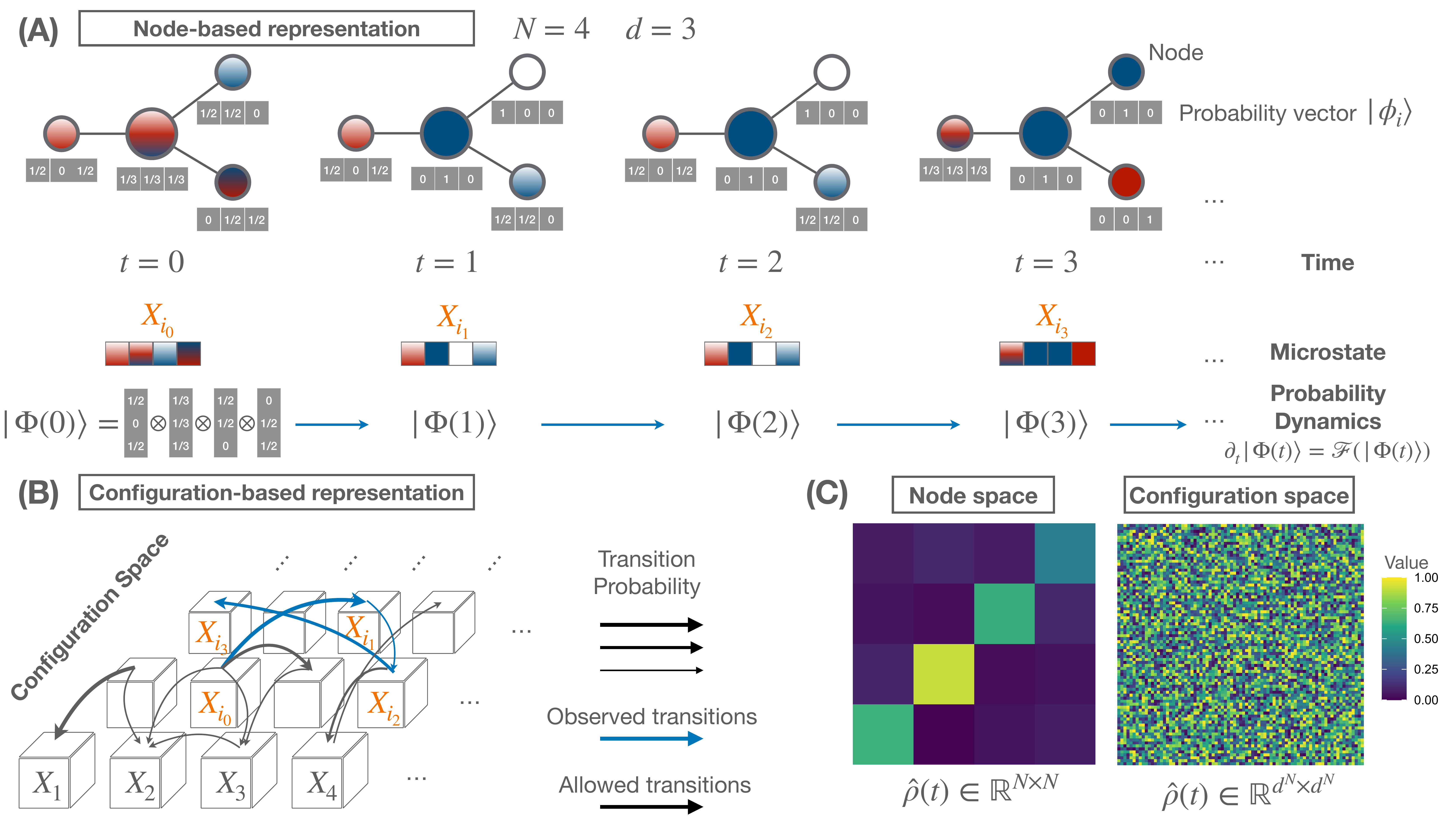}
	\caption{\label{fig:sketch}\textbf{Modeling information dynamics in state-space representation.} (A) An illustrative network consisting of $N=4$ nodes, where each node is characterized by a $d$--dimensional state, with $d=3$. The state $\ket{\Phi(t)}$, that encodes the network microstate obtained from node states, evolves in time according to a master equation (see the text for details). (B) The configuration space of the system consists of all microstates $X$: the systems evolution can be mapped into a trajectory in this space and approximated by transition rules of a Markovian dynamics between the microstates. (C) The density matrix of the system in the node space ($N$ dimensional) and the configuration space ($d^N$ dimensional) are different in size and, as shown later, also in the information they encode about the underlying system. 
 }
\end{figure*}
%
We model the dynamical process on a network stochastically, by supposing each node $X_i$, with $0 < i \leq N$, is a random variable taking values in a discrete and finite set of states $\Sigma$, with $\text{dim}(\Sigma) = d$. The node-based approach assigns to each node a $d$-dimensional probability vector $\ket{\phi_i}$ \footnote{We use the Dirac bra-ket notation to denote vectors $\ket{\phi}$ and their conjugates transpose $\bra{\phi}$}, such that the elements of this vector represent the probabilities $P(X_i = s_i \in \Sigma)$ of the node $i$ taking value in any of the $d$ possible state. Local observables, such as node density or magnetization, can then be computed from $\ket{\phi_i}$, resulting in a $N$-dimensional vector for each possible local observable of interest.  

The configuration-based approach, on the other hand, tracks all $d^N$ possible configurations $X = X_1 X_2 \ldots X_N$ of the entire network (see Fig.~\ref{fig:sketch}). The object of interest is now a probability distribution over all possible network configurations, which may be represented as a $d^N$ dimensional vector in the tensor product basis: 
\begin{widetext}
\begin{equation}\label{manybodyvec}
P\left(X = \prod_{i=1}^N s_i\right) = \ket{\Phi} =  \sum_{s_1, \ldots, s_N \in \Sigma} P(X_1 = s_1; \ldots;  X_N = s_N) \ket{s_1} \otimes \ldots \otimes \ket{s_N}
\end{equation}
\end{widetext}
The node-based representation for $\ket{\phi_i}$ can be obtained from the configuration based representation by marginalizing over all nodes apart from $i$, but the configuration-based approach contains much more information on the system as it tracks not only individual nodes, but also all pairs, triples and any other subset of nodes.
The dynamical evolution of $\ket{\Phi(t)}$ is generally given by 
\begin{equation}
\partial_t \ket{\Phi(t)} = \mathcal{F}(A(t), t, \ket{\Phi(t)}).
\end{equation}
Here, we restrict ourselves to a large class of models which can be suitably described 
by Markovian dynamics on a network with adjacency matrix $A(t)$. This implies imposing the following two assumptions: (i) the change in network state depends linearly on its current state as
\begin{equation}\label{mastereqn}
\partial_t \ket{\Phi(t)} = \hat W(A(t)) \ket{\Phi(t)} \,,
\end{equation}
with $\hat W(A(t))$ the Markov generator in the basis of network configurations; (ii) the dynamics is governed by \emph{local interactions}, i.e., single nodes jump into new states depending only on their current state and the state of its immediate neighbors in the network. 

The last assumption allows us to write the Markov generator $\hat W(A(t))$ as a sum over at most \emph{bilocal} operators, which include operators $\hat c^{i}$ acting on nodes $i$ individually and bilocal operators $\hat a^i \hat b^j$, which act on node $i$ and $j$ simultaneously \footnote{This construction can be extended to higher-order interactions by allowing for $k$-local operators, with $k$ the order of the interaction.}. An explicit construction of the operators $\hat a^i$ is made by inserting a $d$-dimensional matrix operator $\hat a$ at site $i$ into a tensor product with $N-1$ identity matrices. 
Mathematically, the Markov generator $\hat W(A(t))$ can be written as
\begin{equation}\label{genericW}
\hat W(A(t)) = \sum_{\lambda} r_\lambda  \sum_{i,j = 1}^N A^{ij}(t) \hat a^i_{\lambda} \, \hat b^j_{\lambda} + \sum_{\gamma} \sum_{i = 1}^N r_\gamma \hat c_{\gamma}^i,
\end{equation}
where $\lambda$ labels all bilinear (i.e., nearest-neighbor) transitions in the model, each appearing with rate $r_\lambda$, and $\gamma$ denotes the local operators with transition rate $r_\gamma$.  

The conservation and non-negativity of probability imposes constraints on the possible operators which can appear in \eqref{genericW}. By introducing the flat state $\bra{\flat}$, a row vector with all elements equal to 1, it follows that $\bracket{\flat}{\Phi(t)} = 1$, since $\ket{\Phi(t)}$ is a probability vector. Conservation of total probability implies that $\bra{\flat} \hat W = 0$, or: each column of $\hat W$ should sum to zero. In addition, off-diagonal elements of $\hat{W}$ are non-negative to ensure the non-negativity of the probability vector $\ket{\Phi(t)}$ at any $t$. 
Matrices satisfying these properties are \emph{infinitesimally stochastic}, implying that their matrix exponential is a (right-)stochastic matrix. 

Each individual term in \eqref{genericW} should constitute an infinitesimally stochastic operator, which immediately implies that for each $\gamma$, $\hat c_\gamma$ is infinitesimally stochastic. 
The same criteria must hold for the tensor product of matrices $\hat a_{\lambda}$ and $\hat b_{\lambda}$. An explicit basis for all $d(d-1)$ possible infinitesimally stochastic matrices of dimension $d$ can be constructed from the matrices
\begin{equation}\label{qbasis}
\hat q_{k \to l} = \ket{l}\bra{k} - \ket{k}\bra{k} \,, \qquad k,l \in \Sigma.
\end{equation}
For the bilinear interactions, a similar basis is constructed by taking $k,l \in \Sigma \otimes \Sigma$. 

\heading{Density matrix and spectral entropy} For time-independent networks $A(t) = A$, and the solution to the master equation \eqref{mastereqn} is
\begin{equation}
\ket{\Phi(t)} = \hat U(t) \ket{\Phi(0)} \,, \qquad \hat U(t) = \exp\left( \hat W(A) t\right) \,.
\end{equation}
Here $\hat U(t)$ is a stochastic matrix, whose elements correspond to the information flow between network configurations $X$ and $Y$: it is the conditional probability of finding the network in configuration $Y$ at time $t$, given that it started at $t=0$ in configuration $X$: $P(\Phi(t) = Y | \Phi(0) = X) = \bra{Y} \hat U(t) \ket{X}$.

Due to $\bra{\flat}\hat U(t) = \bra{\flat}$, $\ket{\Phi(t)}$ remains a normalized probability distribution at all $t$, and the diagonal elements of $\hat U(t)$ provide the return probabilities for the dynamical process. These are the probabilities that the system is in the same microscopic configuration $X$ after a time $t$, or the amount of information field trapped in the microstate $X$, similar to the node-space representation~\cite{ghavasieh2020statistical}. The total amount of trapped field is the sum over all return probabilities, given by $Z(t) = {\rm Tr}\left[\hat U(t)\right]$. In analogy with quantum mechanics, one can define the density matrix $\hat\rho(t) = \hat U(t)/Z(t)$, such that ${\rm Tr}[\hat\rho(t)] =1$ at all times. 

The spectral entropy $\mathcal{S}(t)$ for the dynamical process is defined as the von Neumann entropy of $\hat{\rho}(t)$:
\begin{equation}\label{spectralS}
\mathcal{S}(t) = - {\rm Tr} \left[ \hat\rho(t) \log \hat\rho(t) \right] \,,
\end{equation}
The spectral entropy is upper-bounded as $\mathcal{S}(t) \leq N \log d$, where the upper bound is saturated when $\hat \rho(t)$ is proportional to the identity matrix. The entropy \eqref{spectralS} is always maximal at $t=0$; at this time the information had no chance to diffuse through the network and all information channels are active by considering all possible initial conditions. At intermediate times, we can write the spectral entropy explicitly as
\begin{equation}\label{SvNexplicit}
\mathcal{S}(t) = - \tr{\hat{W}\hat{\rho}(t)} t + \log Z(t),
\end{equation}
which consists of two manifestly positive terms. The first term is proportional to the rate of change of the trapped field given by $\partial_t \log Z(t)$: it constitutes an average flow, or dissipation, of information out of the trapped field $Z(t)$. The second term is a logarithmic measure of the amount of trapped field $Z(t)$. In analogy to thermodynamics, we interpret the log of the trapped field as internal energy and its dissipation as heat flow. 

At late times $t\to \infty$, $\partial_t Z(t) \to 0$ and the spectral entropy will reduce to $\mathcal{S}(t\to\infty) = \log(n_{\rm s.s.})$, where $n_{\rm s.s.}$ denotes the number of steady-states of the Markov chain defined by $\hat{W}(A)$. As such, $\mathcal{S}(t)$ gives a measure of the availability of different information streams on the space of network configurations at time $t$, ranging from all possible streams at $t=0$, to the minimal number at $t\to \infty$, where it reduces to the (logarithmic) measure of the number of steady states in the system.

\heading{Applications to spin dynamics} It is possible to show that our framework allows to analyze a broad variety of systems and their dynamics. Here we focus on two models whose node-based dynamics are mathematically equivalent, however, on the space of network configurations the dynamics exhibits wildly different behavior: the voter model (VM) and the (symmetric) simple exclusion process (SEP). 

The VM ~\cite{holley1975ergodic} is an emblematic spin system where the node states are $\Sigma = \{+,-\}$. The VM on a network (with link-based update rule) has transitions which align the spin of node $i$ with the spins of neighboring nodes $j$. 
In our framework, these transitions are implemented by 
\begin{equation}\label{Wvoter}
W_{\text{VM}}(A) = \sum_{i,j = 1}^N A^{ij} \left( \hat q_{-\to+}^i \, \hat n_+^j + \hat q_{+\to-}^i  \, \hat n_{-}^j  \right) \,, 
\end{equation}
with matrices $\hat q$ defined in \eqref{qbasis}, $\hat n_+ =  \ket{+}\bra{+}$ and $\hat n_- =  \ket{-}\bra{-}$.
The local magnetization of node $i$ is obtained by projecting the network state vector $\ket{\Phi(t)}$ as
\begin{equation}\label{Mdef}
\vev{m_i} = \bra{\flat} (\hat n_+^i - \hat n_-^i) \ket{\Phi(t)} \,.
\end{equation}
The node-based dynamics of the magnetization, which can be derived from using the master equation \eqref{mastereqn} with \eqref{Wvoter} inside the projection \eqref{Mdef}, is given by:
\begin{equation}\label{GL}
\partial_t \vev{m_i(t)} = - \sum_j L^{ij} \vev{m_j(t)} \,,
\end{equation}
with $L = D -A$ the graph Laplacian and $D$ the diagonal degree matrix, where a nodes degree is the number of its neighbors.

In the simple exclusion process (SEP), each node is either occupied (1) or empty (0). Occupied nodes are thought to contain particles which can hop to neighboring sites, leaving an empty node behind. 
The total number of occupied nodes in the network is hence a conserved quantity. Occupied sites cannot be created or annihilated, but they can move around on the network.
The transition takes neighboring nodes in a product state $\ket{10}$ to the state $\ket{01}$ (and vice-versa), which is implemented by the infinitesimally stochastic matrix
\begin{equation}
\hat q_{10 \to 01} = \ket{01}\bra{10} - \ket{10}\bra{10}\,.
\end{equation}
This transition matrix is not decomposable into a direct product of operators. Instead, it can be written as linear combination of two direct product operators:
\begin{equation}
\hat q_{10 \to 01} = \hat q_{1 \to 0} \otimes \hat n_0 + \hat \sigma_+ \otimes \hat q_{0 \to 1}\,,
\end{equation}
with $\hat q$ matrices defined in \eqref{qbasis}, $\hat n_{0} = \ket{0}\bra{0}$ and $\hat \sigma_{+} = \ket{0}\bra{1}$.
The transition rate matrix for the SEP on a network is hence:
\begin{equation}
W_{\rm SEP}(A) = \sum_{i,j = 1}^N A^{ij} \left( \hat q^i_{1 \to 0}\, \hat n_{0}^j + \hat \sigma_{+}^i \, \hat q_{0 \to 1}^j \right)\,.
\end{equation}
In the case of undirected networks, the local density at node $i$, $\vev{\eta_i(t)} = \bra{\flat} \hat n_1^i \ket{\Phi(t)} $, also evolves according to the graph Laplacian \eqref{GL}. 
Therefore, the node-based dynamics of the magnetization in the VM and the local density in the SEP evolve according to the same equation. In terms of spectral entropies for the node-based dynamics \cite{de2016spectral}, both would be described by the von Neumann entropy of the density matrix $\hat{\rho}(t) = e^{- L t}/Z(t)$ with $Z(t) = {\rm Tr}(e^{-L t})$. 

\begin{figure}[!t]
	\includegraphics[height=0.76\linewidth]{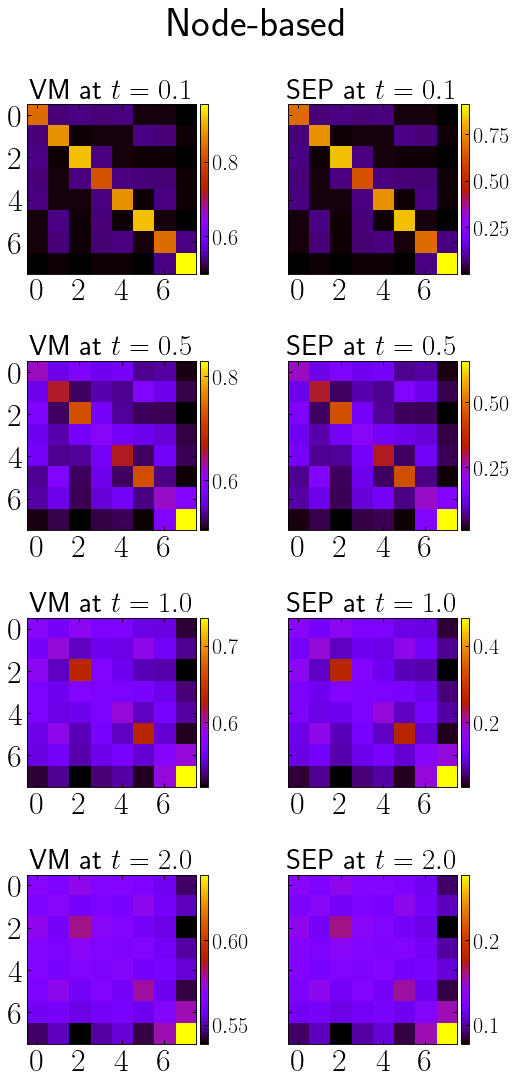}
	\vline
	\includegraphics[height=0.76\linewidth]{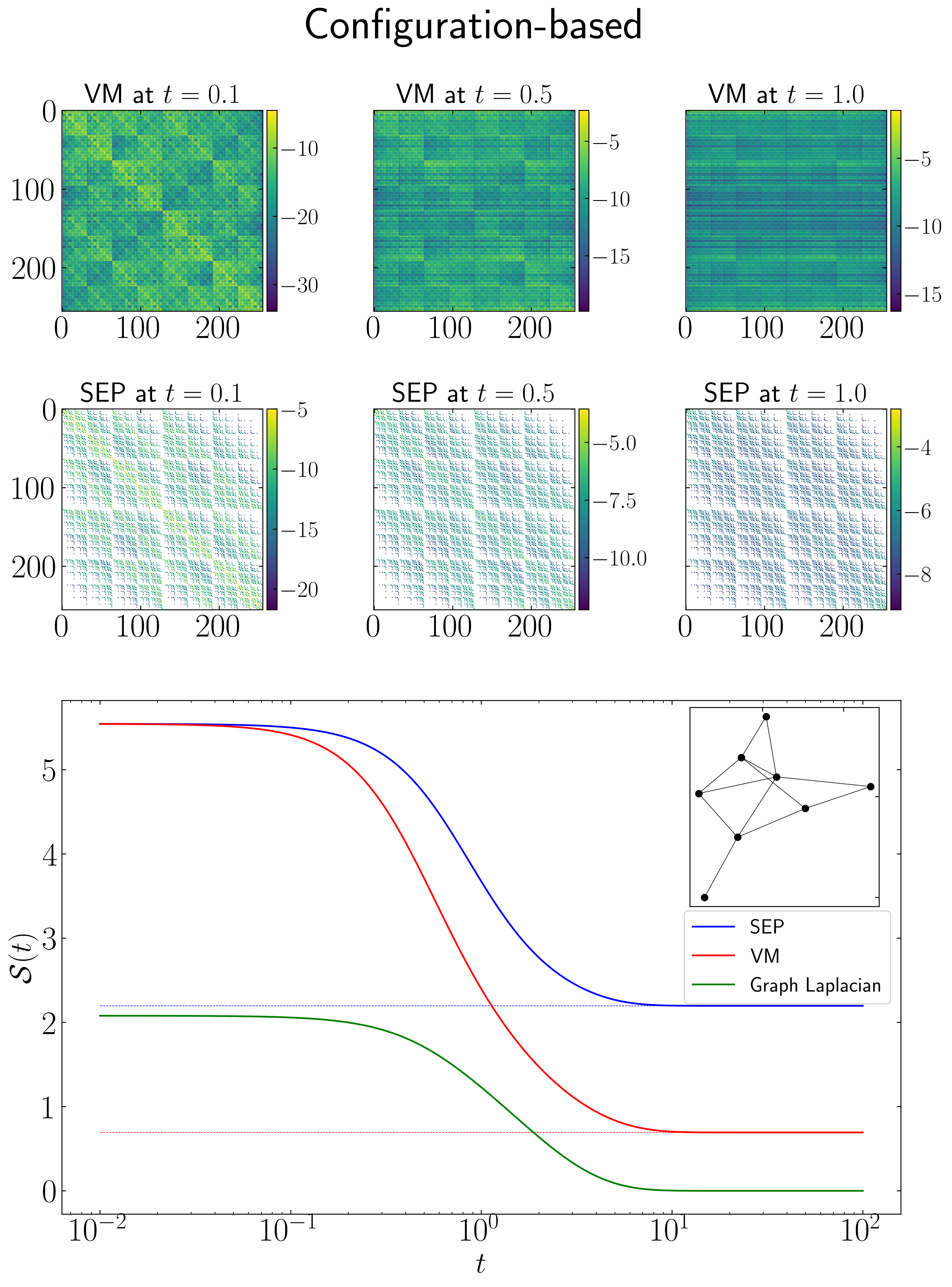}
	\caption{\label{fig:sepvsvot}\textbf{Information dynamics in state-space versus node-space.} Left: 
	The node-based dynamics on a small eight node sample network. Each cell in the heat plot displays the magnetization (VM) or density (SEP) of the node indexed by the row labels at a time $t$ indicated above the plot, conditioned on the event that the node indexed by the column label had spin up (VM) or was occupied (SEP) at $t=0$.  Right top: 
	The configuration-based dynamics for the VM and SEP on the same network. The heat maps show the logarithm of the 
	$2^8$-dimensional density matrix, 
	at three different times. The columns of the plot label the initial configuration $X$ (as an integer whose binary representation corresponds to the network configuration), while the rows label the microstates $Y$ at time $t$ indicated above the plot. White cells means there is no information flow between those microstates. Right bottom: The von Neumann entropy for the two systems as a function of $t$, as compared to the spectral entropy of the graph Laplacian. The inlay shows the network on which this was evaluated. 
	}
\end{figure}

However, the two dynamical processes are not equivalent from a physical point of view. 
The SEP has conserved quantities (total occupancy level) and satisfies local detailed balance, whereas the VM has absorbing states (the two ferromagnetic states with all spins aligned) and breaks local detailed balance. These differences become apparent in the configuration-based framework, as illustrated in Figure~\ref{fig:sepvsvot} for a simple eight node sample network.
The left panel displays heat maps of the node-based dynamics for $\vev{m_i(t)}$ in the VM and $\vev{\eta_i(t)} $ for the SEP. 
Although the absolute values in the heat plots differ, their node-based dynamical evolution is equivalent, as they are both governed by the graph Laplacian.    

The right panel of the Figure~\ref{fig:sepvsvot} shows the evolution for both models on the space of all network configurations. The heat map shows (the logarithm of) the density matrix, quantifying the information flow between all $2^8 = 256$ possible configurations of the network. 
In case of the VM, all microstates -- except for the two absorbing states -- are connected to each other, while the SEP has only certain transitions with non-zero elements. This is due to the fact that each of the occupancy levels in the SEP corresponds to one closed communicating class of the Markov chain. We also show the von Neumann entropy $\mathcal{S}(t)$ of the density matrices $\hat{\rho}(t)$ for this sample network, which at $t=0$ is maximal at the value of $N \log(2)$, as expected. For the VM, the entropy drops to $\log(2)$ at late time, reflecting the two ferromagnetic absorbing states in the system. For the SEP, the entropy drops to $\log(N+1)$, since there are $N+1$ communicating classes in the Markov chain, one for each global occupancy level. The spectral entropy for the node-based dynamics is equivalent to that of the graph Laplacian, which is also shown for reference. 

\begin{figure}
	\includegraphics[width=\columnwidth]{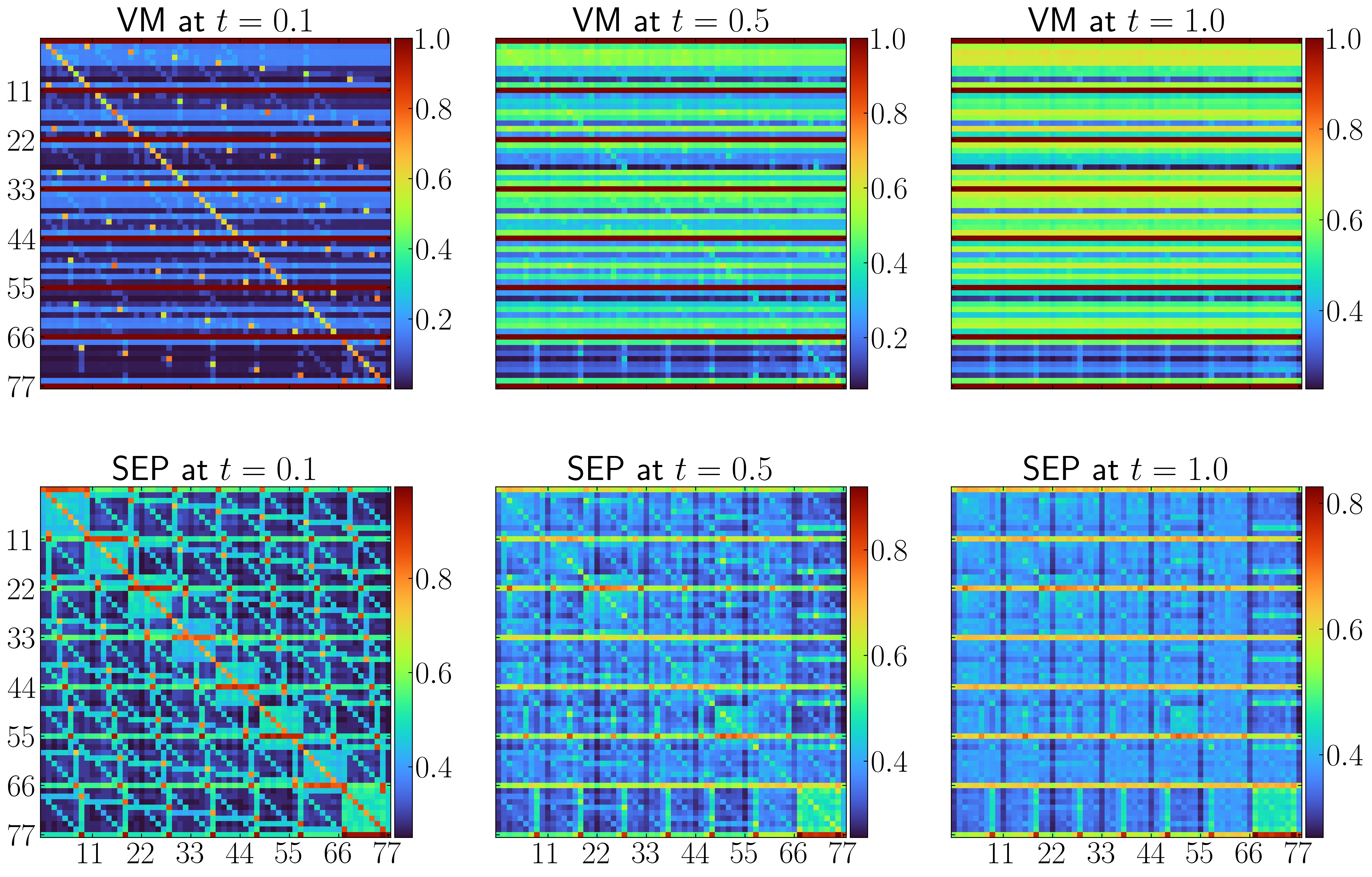}
	\caption{\label{fig:correlators}\textbf{Equal-time correlation functions in VM versus SEP}. The top panels show the magnetization correlations $\langle m_i m_j(t)\rangle $ at different times, where the node pairs $ij$ mark the rows of the heat plots, given that at time $t=0$ node pairs $kl$ (marking the columns) have magnetization $+1$. The bottom panels show similar plots for the density correlations in the SEP, $\vev{\eta_i \eta_j(t)}$, where the initial conditions for the columns marked as $kl$ are such that $\eta_k\eta_l = 1$ at $t=0$.}
\end{figure}

A different projection of the exact $2^N$ dimensional propagator $\hat{U}(t)$ is displayed in Figure~\ref{fig:correlators}, showing the $8^2 \times 8^2$ dimensional matrix of magnetization and density correlation functions $\langle m_i m_j(t) \rangle$ and $\langle \eta_i \eta_j (t)\rangle$. 
These correlations evolve in very different ways for the two dynamical processes, even though the underlying node-based dynamics is equivalent. Physically, this implies that the dynamics of variances in magnetization (VM) and density (SEP) are not equivalent, while the expectation values do evolve in the same way. This would not have been evident from the node-based representation alone.

In conclusion, we have introduced a framework to analyze information flow between network configurations, for a broad class of models where nodes are characterized by vector-valued information fields. 
We show that the spectral entropy of the dynamical system gives a measure of the diversity of available information channels, consisting out of two terms, which represent the amount of information trapped in the set of all network configurations (i.e., internal energy), and the flow of information out of this trapped field (i.e., heat flow). At variance with the spectral entropy in the nodes space, our framework allows to distinguish the behavior of dynamical systems governed by the same (node-based) evolution operator. It is worth remarking that our framework does not lead to a mathematical analytical breakthrough in the treatment of the voter model and the simple exclusion process. Instead, it allows us to analytically demonstrate that those processes, and potentially many other dynamical systems, are dramatically different when a network state-based representation is used. Such an emergent difference would remain hidden when node-based representation is used, as it is the case for state-of-the-art analytical approaches.

We have focused here on exact numerical diagonalization, feasible only for small networks, to show that an exact method leads to a result that has no counterpart when existing methods are used, thus requiring the need for going beyond node-based spectral entropies.
Since the computational complexity of exact diagonalization grows exponentially with system size, for larger networks it is desirable to use approximation techniques that can be borrowed from other research areas. In quantum many-body systems a similar problem occurs, where wavefunctions of composite quantum systems are vectors in the tensor product Hilbert space of exponentially large complexity.
However, whenever the system is dominated by local interaction, powerful mathematical and numerical methods \cite{white1992density,schollwock2011density,orus2014practical} have been developed to find efficient and accurate representation for the many-body wavefunction using tensor networks. 
Recent work has shown applications of tensor networks to lattice models of stochastic systems  \cite{helms2019dynamical,banuls2019using,causer2021optimal,strand2022using,garrahan2022topological,causer2022finite}.
An interesting question is whether and for which classes of network models efficient and accurate representations in terms of tensor networks exist. 

\begin{acknowledgments}
MDD acknowledges financial support from the Human Frontier Science Program Organization (HFSP Ref. RGY0064/2022), from the University of Padua (PRD-BIRD 2022) and from the EU funding within the MUR PNRR “National Center for HPC, BIG DATA AND QUANTUM COMPUTING” (Project no. CN00000013 CN1).
\end{acknowledgments}


%

\end{document}